\documentclass[prl,aps,twocolumn,epsfig]{revtex4}
\usepackage{graphicx}
\begin{document}
\title{Roton-maxon spectrum and stability of trapped dipolar condensates}
\author{L. Santos$^{1}$, G. V. Shlyapnikov$^{1,2,3}$, and
M. Lewenstein$^{1}$}
\address{(1)Institut f\"ur Theoretische Physik, Universit\"at Hannover,
 D-30167 Hannover,
Germany\\ 
(2) FOM Institute for Atomic and Molecular Physics,
Kruislaan 407, 1098 SJ Amsterdam, The Netherlands\\
(3) Russian Research Center Kurchatov Institute, Kurchatov Square, 123182
 Moscow, Russia
}
\begin{abstract}
We find that pancake dipolar condensates can exhibit a roton-maxon character of
the excitation spectrum, so far only observed in superfluid helium. We also
obtain a condition for the dynamical stability of these condensates. The
spectrum and the border of instability are tunable by varying the particle
density and/or the confining potential. This opens wide possibilities for
manipulating the superfluid properties of dipolar condensates. 
\end{abstract} 
\maketitle


Recent progress in cooling and trapping of polar molecules \cite{John,Gerard}
opens fascinating prospects for achieving quantum degeneracy in trapped
gases of dipolar particles. Being electrically or magnetically polarized,
polar molecules interact with each other via long-range anisotropic
dipole-dipole forces. This makes the properties of such dipolar gases
drastically different from the properties of commonly studied atomic cold
gases, where the interparticle interaction is short-range. Other candidates to
form a dipolar gas are atoms with large magnetic moments
\cite{dipolar_atoms,Goral}, and atoms with dc-field-\cite{You} or light-induced
electric dipole moments  \cite{Giovanazzi,Santos,DDgates}.  

The dipole-dipole interaction is responsible for a variety of novel phenomena
in ultracold dipolar systems. The
energy independence of the dipole-dipole scattering amplitude for any orbital
angular momenta provides realistic possibilities for achieving a superfluid BCS
transition in single-component dipolar Fermi gases (see \cite{Baranov} and
refs. therein). Dipolar bosons in optical lattices have been shown to provide
a highly controllable environment for engineering various quantum phases 
\cite{Gorallattice}. In addition to superfluid and Mott-Insulator ones,
recently observed in Munich experiments \cite{Greiner} with bosonic atoms, 
the long-range dipole-dipole potential provides
supersolid and checker-board phases. Dipole-dipole interactions are also 
responsible for spontaneous polarization and spin waves in spinor condensates 
in optical lattices \cite{Meystre}, and may lead to self-bound
structures in the field of a traveling wave \cite{Giovanazzi2}.
Recently, dipolar particles have been considered as promising candidates for
the implementation of fast and robust quantum-computing schemes
\cite{DDgates,qcomp}.

The long-range and anisotropic (partially attractive) character of dipole-dipole 
forces ensures a strong dependence of the stability
of trapped dipolar Bose-Einstein condensates on the trapping geometry
\cite{Santos}. For cylindrical traps with the aspect ratio
$l_z/l_{\rho}>l_*=0.43$, a purely dipolar condensate is dynamically unstable
if the number of particles $N$ exceeds a critical value. A detailed study of
the excitation modes for this geometry is contained in Ref. \cite{Goralexcit}.
It has also been argued in Ref. \cite{Santos} that in pancake traps with 
$l_z/l_{\rho}<l_*$ the ground state solution is expected for any $N$.

In this Letter we analyze the nature of excitations and instability of 
pancake-shaped dipolar condensates. For this purpose, we consider the 
physically transparent case of an infinite pancake trap, with the dipoles
perpendicular to the trap plane. For the maximum condensate
density $n_0\rightarrow\infty$, the dynamical stability requires the presence
of a sufficiently large short-range repulsion, in addition to the
dipole-dipole interaction. Otherwise, if $n_0$ exceeds a critical value
$n_c$, excitations with certain (large) in-plane momenta $q$ become
unstable. At densities $n<n_c$ the excitation spectrum has a roton-maxon
form (see Fig.1) similar to that in superfluid helium. 

The roton-maxon spectrum for helium has been suggested by Landau \cite{Landau},
and later Feynman \cite{Feynman} related the excitation energy to the static
structure factor of the liquid. The roton minimum originates from the fact
that at intermediate momenta one has a local structure produced by the
tendency of atoms to stay apart. The pancake dipolar condensate is the first 
example of a weakly interacting gas, where the spectrum has a roton-maxon form. 
In an infinite pancake trap this spectrum allows a transparent physical 
interpretation. 

We emphasize that the roton-maxon spectrum finds its origin in the momentum
dependence of the interparticle interaction. For in-plane
momenta $q$ much smaller than the inverse size $L$ of the condensate in the
confined direction, excitations have two-dimensional (2D) character. Hence,
as the dipoles are perpendicular to the plane of the trap, particles
efficiently repel each other and the in-plane excitations are phonons.  
For $q\gg 1/L$, excitations acquire 3D character and the interparticle
repulsion is reduced. This decreases the excitation energy under an increase of
$q$. The energy reaches a minimum and then starts to grow as the excitations
continuously enter the single-particle regime. The minimum energy is zero for
the maximum density $n_0$ equal to a critical value. At higher
densities, excitations with momenta $q$ in the vicinity of this 
minimum become unstable and the condensate collapses. Below we find the
condensate wave function, excitation spectrum, and the conditions for both
rotonization and instability.  

We consider a condensate of dipolar particles harmonically confined in the
direction of the dipoles ($z$) and uniform in two other directions
($\mbox{\boldmath$\rho$}=\{x,y\}$). The dynamics of the condensate wave 
function $\psi({\bf r},t)$ in this  infinite pancake trap is described by 
the time-dependent Gross-Pitaevskii (GP) equation (see \cite{Santos} and 
refs. therein):  
\begin{eqnarray} && 
i\hbar\frac{\partial}{\partial t}\psi({\bf r},t)=\left \{
-\frac{\hbar^2}{2m}\Delta+\frac{m}{2}\omega^2z^2+g|\psi({\bf r},t)|^2
\right \delimiter 0 \nonumber \\ && \left \delimiter 0 
+d^2\int d{\bf r}'V_d({\bf r}-{\bf r}')
|\psi({\bf r}',t)|^2 \right \}\psi({\bf r},t), 
\label{GPE}
\end{eqnarray}
where $\omega$ is the confinement frequency, $m$ is the particle mass, and $d$
the dipole moment. The wave function $\psi({\bf r},t)$ is normalized to the
total number of particles. The third term in the rhs of Eq.(\ref{GPE})
corresponds to the mean field of short-range (Van der Waals) forces, and the
last  term to the mean field of the dipole-dipole interaction. The coupling
constant for the short-range interaction is $g$, and $V_d(\vec
r)=(1-3\cos^2\theta)/r^3$ is the potential of the dipole-dipole interaction,
with $\theta$ being the angle between the vector $\vec{r}$ and the direction
of the dipoles ($z$).
 
The ground state wave function is independent of the in-plane coordinate 
$\mbox{\boldmath$\rho$}$ and can be written as $\psi_0(z)\exp{(-i\mu t)}$,
where $\mu$ is the chemical potential. Then, integrating over
$d\mbox{\boldmath$\rho$}\,'$ in the dipole-dipole term of Eq.(\ref{GPE}), we
obtain a one-dimensional equation similar to the GP equation for short-range
interactions:  
\begin{equation}
\!\!\left \{ -\frac{\hbar^2}{2m}\Delta+\frac{m}{2}\omega^2z^2+(g+g_{d})
\psi_0^2(z)\!-\!\mu\right \}\psi_0(z)=0,
\label{GPE1D}
\end{equation}
where $g_{d}=8\pi d^2/3$. We will discuss the case of $(g+g_d)>0$, where the
chemical potential $\mu$ is always positive. For $\mu\gg\hbar\omega$ the
condensate presents a Thomas-Fermi (TF) density profile in the confined
direction: $\psi_0^2(z)= n_0(1-z^2/L^2)$, with $n_0=\mu/(g+g_d)$ being the
maximum density, and $L=(2\mu/m\omega^2)^{1/2}$ the TF size.   

Linearizing Eq.(\ref{GPE}) around the ground state solution $\psi_0(z)$ we
obtain the Bogoliubov-de Gennes (BdG) equations for the excitations. Those
are characterized by the momentum ${\bf q}$ of the in-plane free motion and by
an integer quantum number ($j\geq 0$) related to the motion in the $z$
direction. The excitation wave functions take the form $f_{\pm}(z)\exp(i{\bf
q}\mbox{\boldmath$\rho$})$,  where $f_{\pm}=u\pm v$, and $u,v$ are the
Bogoliubov $\{u,v\}$ functions. Then the BdG equations read:
\begin{mathletters} \begin{eqnarray} 
\epsilon f_{-}&=&\frac{\hbar^2}{2m} \left [
-\frac{d^2}{dz^2}+q^2+\frac{\Delta\psi_0}{\psi_0}\right ] f_{+}
\equiv H_{kin}f_+, 
\label{BdGa} \\
\epsilon f_{+}&=&H_{kin}f_{-}+H_{int}[f_-],  \label{BdGb} 
\end{eqnarray} \end{mathletters} 
where $H_{kin}$ is the sum of kinetic energy operators, and
\begin{eqnarray}
& & H_{int}[f_-]=2(g_{d}+g)f_{-}(z)\psi_0^2(z)-(3/2)g_d\,q\,
\psi_0(z)\times \nonumber \\ 
& &
\times\int_{-\infty}^{\infty}
d z'\,f_{-}(z')\psi_0(z')\exp{(-q|z-z'|)}.  
\label{Hint}
\end{eqnarray}
For each $j$ we get the excitation energy
$\epsilon_j$ as a function of $q$. We will be mostly interested in the lowest
excitation branch $\epsilon_0(q)$ for which the confined motion is not excited
in the limit $q\rightarrow 0$. 

The second term in the rhs of Eq.(\ref{Hint}) originates from the non-local
character of the dipole-dipole interaction and gives rise to the momentum
dependence of an effective coupling strength. In the limit of low in-plane
momenta $qL\ll 1$, this term can be omitted. In this case, excitations of the
lowest branch are essentially 2D and the effective coupling
strength corresponds to repulsion. Equations (\ref{BdGa}) and  (\ref{BdGb})
become identical to the BdG equations for the excitations of a trapped
condensate with a short-range interaction characterized by a
coupling constant $(g+g_d)>0$. In the TF regime for the confined motion, the
spectrum of low-energy excitations for this case has been found by Stringari
\cite{Stringari}. The lowest branch represents phonons propagating in the
$x,y$-plane. The dispersion law and the sound velocity $c_s$ are given by 
\begin{equation}     \label{phonon}
\epsilon_0(q)=\hbar c_sq;\,\,\,\,\,\,\,c_s=(2\mu/3m)^{1/2}.    
\end{equation}

For $qL\gg 1$, the excitations become 3D and the effective coupling strength
decreases. The interaction term is then reduced to
$H_{int}[f_-]=(2g-g_d)\psi_0^2(z)f_-(z)$ as in the integrand of
Eq.~(\ref{Hint}) one can put $z'=z$ in the arguments of $f_-$ and $\psi_0$. In
this case, Eqs.~(\ref{BdGa}) and (\ref{BdGb}) are similar to the BdG equations
for the excitations of a condensate with short-range interactions characterized
by a coupling constant $(2g-g_d)$. If the parameter $\beta=g/g_d>1/2$, this
coupling constant is positive and one has excitation energies which are real
and positive for any momentum $q$ and condensate density $n_0$. For
$\beta<1/2$, the coupling constant is negative and one easily shows that at
sufficiently large density the condensate becomes unstable. For collective 
excitations  in the TF regime at $n_0\rightarrow\infty$, kinetic energy terms
in  Eq.(\ref{BdGb}) can be omitted and it reads $\epsilon
f_+=(2g-g_d)n_0(z)f_-$. Then, rescaling the excitation  energies as
$\epsilon^2=\tilde\epsilon^2(2g-g_d)/(g+g_d)$, Eqs.~(\ref{BdGa}) and
(\ref{BdGb}) give the eigenmode equation  $\tilde\epsilon^2
f_+=2(g+g_d)n_0(z)H_{kin}f_+$ for positive excitation energies
$\tilde\epsilon$ in the case of short-range repulsive interaction with the
coupling constant $(g+g_d)$. Thus, for $\beta<1/2$ we obtain $\epsilon^2<0$
and imaginary $\epsilon$, which indicates dynamical instability of the
condensate with regard to these high-momentum excitations.   

We thus see that the most interesting behavior of the excitation spectrum in
the TF regime is expected for $qL\gg 1$ and $\beta$ close to the critical value 
$1/2$. In our analytical analysis we first  reduce Eqs.~(\ref{BdGa}) and
(\ref{BdGb}) to the equation for the function $W$ defined by the
relation $f_+\!=\!W\sqrt{1-x^2}$, where $x\!=\!z/L$. Expressing the function
$f_-$ through $W$ from Eq.(\ref{BdGa}),  we substitute it into Eq.(\ref{BdGb})
and integrate straightforwardly over  $dz'$ in $H_{int}[f_-]$  as the main
contribution to the integral comes from a narrow range of distances
$|z'-z|\sim 1/q$. This yields   
\begin{eqnarray}      
\left[\frac{1}{2}(1-x^2)\frac{d^2W}{dx^2}-\left(1+\frac{3}{2(1+\beta)}
\right)x\frac{dW}{dx}\right]\hbar^2\omega^2+     \nonumber \\  
\left[\epsilon^2-E_q^2-\frac{2\beta-1}{1+\beta}\mu
E_q(1-x^2)-\frac{3\hbar^2\omega^2}{2(1+\beta)}\right]W=0,   
\label{eigenmode}
\end{eqnarray}   
where $E_q=\hbar^2q^2/2m$. Here we omitted terms of the order
of $E_q\hbar^2\omega^2/\mu$ and $\hbar^4\omega^4/\mu^2$, since they
are small compared to either $\hbar^2\omega^2$ or $E_q^2$.

For each mode of the confined motion (each quantum number $j$), the
solution of Eq.(\ref{eigenmode}) can be written as series of expansion in
Gegenbauer polynomials $C^{\lambda}_n(x)$, where
$\lambda=(4+\beta)/2(1+\beta)$, and $n\geq 0$ is an integer. The coupling
between polynomials of different power is provided by the term proportional to
$(2\beta -1)(1-x^2)W$. For the critical value $\beta=1/2$ the coupling is
absent, and we then obtain $W_j\propto C^{\lambda}_j(x)$. The dispersion
law is characterized by a plateau (see Fig.1a), and for the $j$-th branch
of the spectrum it is given by  
\begin{equation}    \label{jb}
\epsilon^2_j(q)=E_q^2+\hbar^2\omega^2(1+j(j+3)/2);\,\,\,\beta=1/2,\,\,qL\gg 1.
\end{equation}

For $\beta\neq 1/2$, assuming that the coupling term $\mu
E_q|2\beta-1|/(1+\beta)\alt\hbar^2\omega^2$ and it does not significantly
modify the eigenfunctions, we can still confine ourselves to the 
perturbative approach. 
Then, as the polynomials $C^{\lambda}_j$ are orthogonal with the
weight $(1-x^2)^{\lambda-1/2}$, for the lowest branch of the spectrum we obtain
\begin{equation}     \label{jbb}
\epsilon^2(q)=E_q^2+\frac{(2\beta-1)(5+2\beta)}{3(1+\beta)(2+\beta)}E_q\mu
+\hbar^2\omega^2;\,\,\,\,qL\gg 1.
\label{esp}
\end{equation}
From Eq.(\ref{jbb}) one sees two types of behavior of the spectrum. For
$\beta>1/2$ the excitation energy monotonously increases with $q$ (see Fig.1b).
If $\beta<1/2$, then the dispersion law (\ref{jbb}) is characterized by the
presence of a minimum. Since in the limit of $qL\ll 1$ the
energy $\epsilon_0$ grows with $q$, the existence of this minimum indicates
that the spectrum as a whole should have a roton-maxon character (see Fig.1b).
This behavior is known from the physics of liquid helium. As discussed above,
in our case it is related to the reduction in the coupling
strength with an increase of momentum, resulting from the transformation of
the character of excitations from 2D to 3D. 

As follows from Eq.(\ref{jbb}) for $\beta$ close to $1/2$, the roton minimum
is located at $q=(16\mu\delta/15\hbar\omega)^{1/2}1/l_0$, where
$\delta=1/2-\beta$, and $l_0=(\hbar/m\omega)^{1/2}$ is the harmonic oscillator
length for the confined motion. The excitation energy at this point is 
$\epsilon_{\rm min}=[\hbar^2\omega^2-(8\mu\delta/15)^2]^{1/2}$. An increase of
the density (chemical potential) or $\delta$ makes the roton
minimum deeper. For $\mu\delta/\hbar\omega=15/8$ the minimum energy reaches
zero at $q=\sqrt{2}/l_0$. At larger values of $\mu\delta/\hbar\omega$ one gets
imaginary excitation energies for $q\sim 1/l_0$, and the condensate becomes
unstable.

We have then found the excitation spectrum numerically from Eqs.
(\ref{BdGa}) and (\ref{BdGb}) for various values of $\beta$ and
$\mu/\hbar\omega$. The results for the TF regime and $\beta$ close to $1/2$
are presented in Fig.1, where one sees a good agreement between the numerics
and analytics. The discrepancy is mainly due to the neglect of the border
effects and some of the kinetic energy terms when obtaining
Eq.~(\ref{eigenmode}) from Eqs.~(\ref{BdGa}) and (\ref{BdGb}). A similar
behavior of the spectrum is observed for non-TF condensates. In this case, due
to a large kinetic energy in the  confined  direction, the stability of the
condensate does not require as strong a short-range repulsive coupling
strength as in the TF regime. Accordingly, the rotonization of the spectrum
and the instability appear at smaller values of $\beta$. These critical
$\beta$ have been calculated numerically as functions of $\mu/\hbar\omega$ and
are shown in Fig.2. 

The dipolar condensate is the first example of a weakly interacting gas
offering a possibility of obtaining a roton-maxon dispersion,  up to now only
observed in the relatively more complicated physics of liquid He. In contrast
to the helium case, the rotonization in dipolar condensates is 
{\em tunable}. By
varying the density, the frequency of the tight confinement, and the
short-range coupling one can manipulate and control the spectrum,
making the roton minimum deeper or shallower. One can also eliminate it
completely and get the Bogoliubov-type spectrum or, on the opposite, reach the
point of instability. 

\begin{figure}
\includegraphics[width=5.5cm,height=5.0cm]{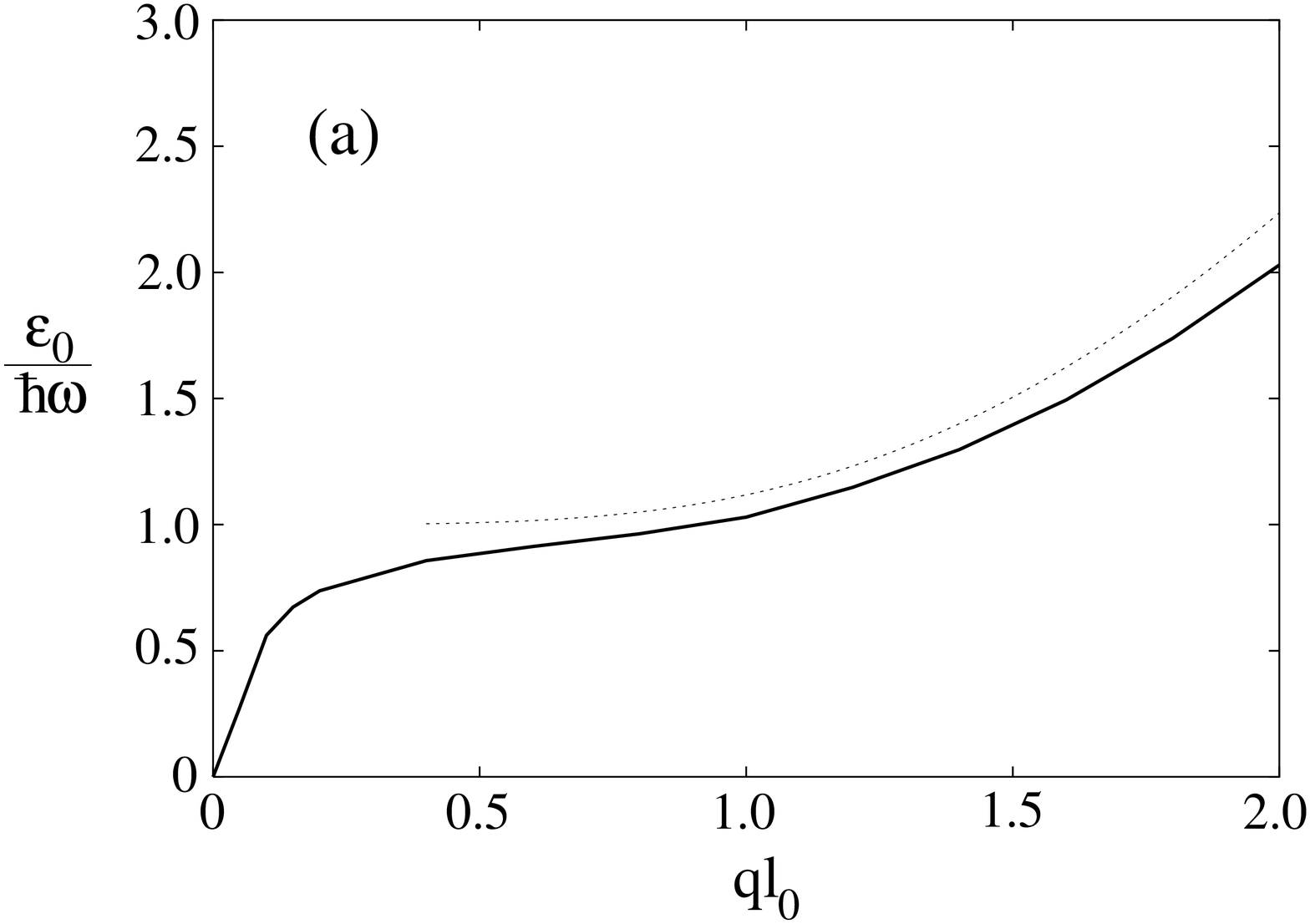}
\includegraphics[width=5.5cm,height=5.0cm]{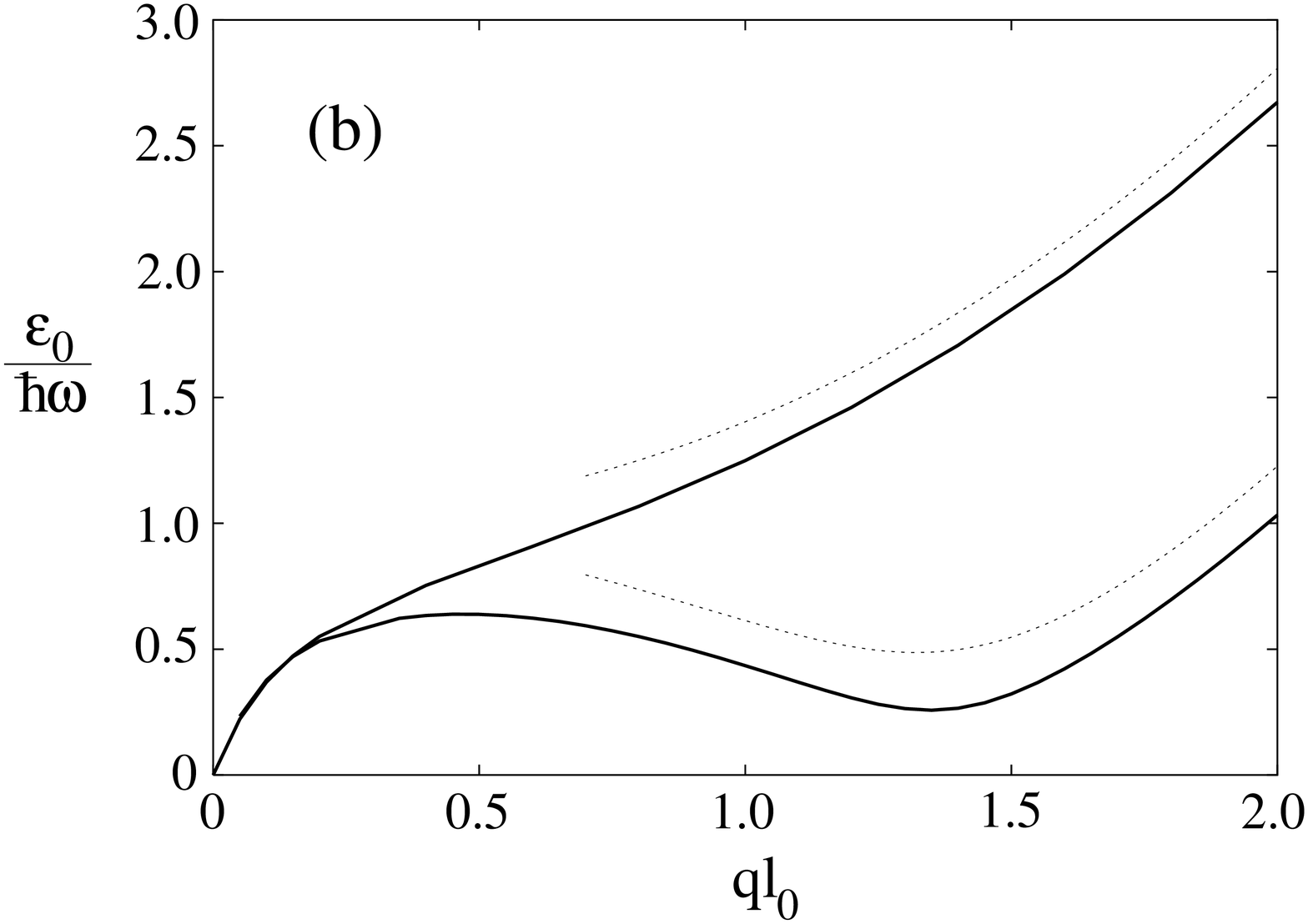} 
\caption{Dispersion law $\epsilon_0(q)$ for various values of $\beta$ and 
$\mu/\hbar\omega$: (a) $\beta=1/2$, $\mu/\hbar\omega=343$; 
(b) $\beta=0.53$, $\mu/\hbar\omega=46$ (upper curve) 
and $\beta=0.47$, $\mu/\hbar\omega=54$ (lower curve). 
The solid curves show the numerical results, and the dotted curves the result 
of Eq.~(\ref{esp}).}
\label{fig:1} 
\end{figure}

\begin{figure}
\includegraphics[width=7.2cm,height=5.5cm]{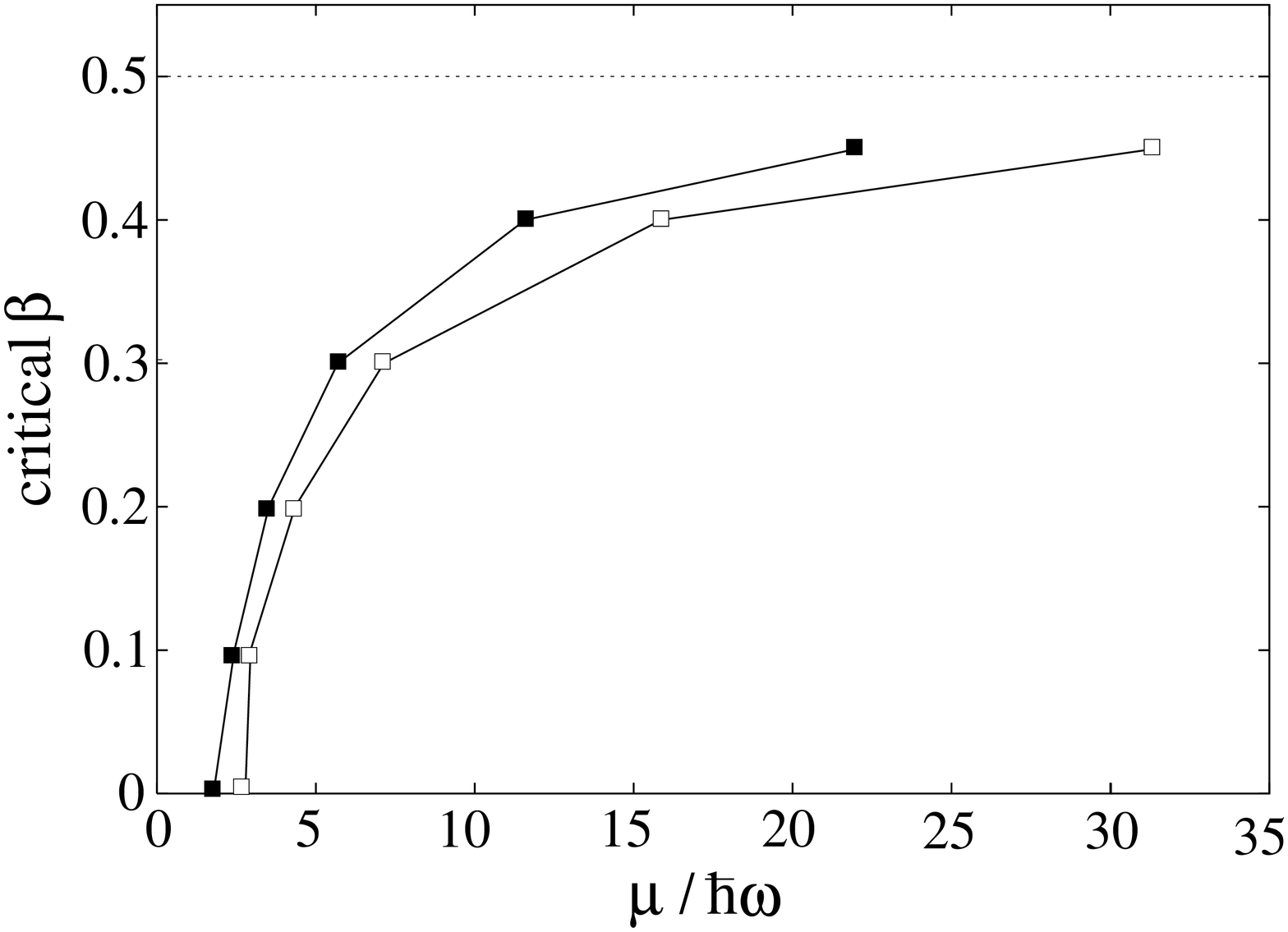}
\caption{Critical values of $\beta$ for the rotonization (filled squares) 
and for the instability (hollow squares) versus $\mu/\hbar\omega$.}
\label{fig:3} 
\end{figure}

The instability of dipolar condensates with regard to short-wave
excitations, is fundamentally different from the well-known instability of
condensates with attractive short-range interaction (negative scattering
length). In the latter case the chemical potential is negative and the ground
state does not exist. The unstable excitations are long-wave and an
infinitely large cloud undergoes local collapses. For the dipolar BEC the
chemical potential is positive and the instability is related to the momentum
dependence of an effective coupling strength. The unstable excitations become
the ones with high momenta at which the coupling is attractive. The existence
of the roton minimum at a given $\beta<1/2$ for $\mu/\hbar\omega$ just below
the point of instability, is likely to indicate that there is a new ground
state in the region of the condensate instability. The presence and character
of this state will be a subject of our future studies.

The presence of the roton minimum in the excitation spectrum can be observed
in Bragg-spectroscopy experiments as those of Steinhauer {\it et al.} 
\cite{Ozeri}, or
in the MIT-type of measurement of the critical velocity for
superfluidity \cite{Ketterle}. According to the Landau criterion \cite{LL}, 
the critical velocity $v_c$ is equal to the minimum value of $\epsilon_0(q)/q$,
and the presence of the roton minimum strongly reduces $v_c$. Even in the
absence of rotonization, a decrease in the slope of the dispersion curve at 
large momenta leads to a significant reduction of the critical velocity.

In conclusion, we have found that pancake dipolar condensates can exhibit 
a roton-maxon character of the excitation spectrum. The presence, position,
and depth of the roton minimum are tunable by varying the density, confining
potential, and the short-range coupling strength. This opens new handles on
manipulations of superfluid properties of trapped condensates.  

We acknowledge support from Deutsche Forschungsgemeinschaft,  
the Alexander von Humboldt Foundation, the RTN Cold Quantum Gases, ESF PESC
BEC2000+, the Dutch Foundations FOM and NWO, INTAS, and the
Russian Foundation for Fundamental Research. L. S. wishes to thank the 
Federal Ministry of Education and  Research and the ZIP
Programme of the German Government.

Note added: After this work has been completed, we 
become aware of 
the recent preprint 
by O'Dell {\it et al.} \cite{ODell} 
where a roton-maxon spectrum has been obtained 
numerically for some particular cases in elongated condensates with 
laser-induced dipole-dipole interatomic interactions.

\end{document}